\documentstyle[psfig]{l-aa}
\begin{document}

\thesaurus{02.01.2, 08.09.2 (GU Mus, GS 1124-68), 13.25.5 --- Section 06}

\title{Spectrophotometric follow--up of GU Mus, the (X--ray) Nova Muscae 
1991$^{\star}$}
\thanks{Based on observations obtained at the European Southern Observatory, 
La Silla, Chile.}

\author{M. Della Valle\inst{1}, N. Masetti\inst{1} \& A. Bianchini\inst{1}}

\institute{(1) Dipartimento di Astronomia, Universit\`a di Padova, 
vicolo dell'Osservatorio, 5 I-35122 Padua, Italy}
\offprints{Nicola Masetti, masetti@astrpd.pd.astro.it}

\date{Received February 1997; accepted 25 August 1997}

\maketitle
\markboth{M. Della Valle et al.: Spectrophotometry of GU Mus}{}

\begin{abstract}

The spectrophotometric follow--up of the X--ray nova GU Mus (=GS 1124-68) is
presented. We have analysed 128 $B$, $V$ and $R$ frames and 19 optical spectra, 
collected at ESO--La Silla, from the epoch of the discovery to January 1992. 
The optical lightcurve appears to be typical for a X--ray nova and exhibits 
secondary maxima in coincidence with the X--ray ones. The spectra show that 
Balmer and He {\sc ii} $\lambda$4686 emission lines and the N {\sc iii} 
$\lambda$4640 emission blend widen during the decline. A possible explanation 
of this behaviour is proposed and applied to the spectral evolution 
of other X--ray novae.


\end{abstract}

\section{Introduction}

Soft X--ray Transients (or SXTs) form a small subclass of Low Mass X--ray
Binaries (LMXBs) and are composed of a late type low--mass star (usually a
dwarf or a subgiant) which loses matter via Roche--lobe overflow onto a massive
collapsed primary.
The importance of these systems lies in the fact that about 80$\%$ of them are 
suspected to harbor a black hole. This has been demonstrated
in 6 cases: V616 Mon (McClintock \& Remillard 1986), GU Mus
(Remillard et al. 1992), GRO J1655-40 (Bailyn et al. 1995b), V2107 Oph
(Remillard et al. 1996), QZ Vul (Casares et al. 1995) and V404 Cyg (Casares
et al. 1992), whose mass functions exceed 3 $M_\odot$.

\smallskip
GS 1124-68 (=X--Ray Nova Muscae 1991) was discovered as a bright transient 
X--ray source on January 8, 1991 by the Ginga (Makino 1991) and the 
GRANAT (Lund \& Brandt 1991) satellites.
The analysis of the two--component X--ray spectrum during the outburst 
(Kitamoto et al. 1992, Grebenev et al. 1992, Ebisawa et al. 1994) allowed the 
classification of this object as SXT.

Its optical counterpart, GU Mus, was discovered by Della Valle et al. (1991)
one week later, on January 15, 1991. GU Mus increased its luminosity from 
$V\sim20.5$ to $V\sim13.5$, a brightening of $\sim$7 
mag which is typical of X--ray novae (van Paradijs \& McClintock 1995).
The optical spectrum at maximum (Della Valle et al. 1991) showed Balmer,
He {\sc i}, He {\sc ii}, N {\sc ii} and N {\sc iii} emission features 
superimposed on a blue optical continuum, thus resembling those of outbursting 
Dwarf Novae (DNe).
From these observations a color excess of 0.30 mag was derived, confirmed later
on by the analysis of UV spectra (Cheng et al. 1992).
In addition, Ball et al. (1995) detected a transient radio emission during 
maximum, which is another typical feature of outbursting SXTs.

\begin{table*}
\caption[]{The journal of the observations. The upper part of the Table lists 
the photometric data, while the spectroscopic observations are reported in
the lower part}
\begin{center}
\begin{tabular}{ccccc}
\noalign{\smallskip}
\hline
\noalign{\smallskip}
Date & Telescope & Filter & Number & Exp. times \\
 & & or passband & of frames & (minutes) \\
\noalign{\smallskip}
\hline
\noalign{\smallskip}
\hline
\noalign{\smallskip}
\multicolumn{5}{c}{Imaging} \\
\noalign{\smallskip}
\hline
\noalign{\smallskip}
\multicolumn{1}{l}{Jan. 15, 1991} & 2.2m,NTT & $B,V,R$ & 2,2,3 & 
0.083,0.16,0.25,0.5 \\
\multicolumn{1}{l}{Jan. 16, 1991} & 2.2m & $B,V,R$ & 1,1,2 & 0.16,0.33 \\
\multicolumn{1}{l}{Jan. 17, 1991} & 2.2m & $B,V,R$ & 1,1,1 & 0.16,0.5 \\
\multicolumn{1}{l}{Jan. 18, 1991} & 2.2m & $B,V,R$ & 1,1,1 & 0.25,0.5 \\
\multicolumn{1}{l}{Jan. 19, 1991} & 2.2m & $B,V,R$ & 1,1,1 & 0.42,0.66 \\
\multicolumn{1}{l}{Jan. 20, 1991} & 2.2m & $B,V,R$ & 1,1,1 & 0.42,0.66 \\
\multicolumn{1}{l}{Jan. 21, 1991} & 2.2m & $B,V,R$ & 1,1,1 & 0.42,0.66 \\
\multicolumn{1}{l}{Jan. 22, 1991} & 2.2m & $B,V,R$ & 1,1,1 & 0.42,0.66 \\
\multicolumn{1}{l}{Jan. 23, 1991} & NTT & $V,R$ & 1,1 & 0.16,0.33 \\
\multicolumn{1}{l}{Jan. 24, 1991} & NTT & $V,R$ & 1,2 & 0.2,0.42 \\
\multicolumn{1}{l}{Jan. 29, 1991} & 2.2m & $B,V,R$ & 15,1,1 & 0.5,1 \\
\multicolumn{1}{l}{Feb. 11, 1991} & NTT & $R$ & 2 & 0.083,0.5 \\
\multicolumn{1}{l}{Feb. 17, 1991} & 2.2m & $B,V,R$ & 1,1,3 & 0.16,0.33 \\
\multicolumn{1}{l}{Feb. 18, 1991} & 2.2m & $B,V,R$ & 2,2,3 & 0.16,0.33 \\ 
\multicolumn{1}{l}{Feb. 19, 1991} & 2.2m & $B,V,R$ & 1,1,1 & 0.25,0.5 \\
\multicolumn{1}{l}{Feb. 21, 1991} & 3.6m & $B,V,R$ & 1,1,1 & 0.13,0.25 \\
\multicolumn{1}{l}{Feb. 24, 1991} & Danish & $B,V,R$ & 3,1,2 & 
0.25,0.33,0.5,5,10 \\
\multicolumn{1}{l}{Feb. 28, 1991} & NTT & $B,V,R$ & 2,1,1 & 0.1,0.15,0.33 \\
\multicolumn{1}{l}{Mar. 1, 1991} & NTT & $B,V,R$ & 1,1,1 & 0.1,0.33 \\
\multicolumn{1}{l}{Mar. 2, 1991} & NTT & $B,V,R$ & 1,1,1 & 0.1,0.33 \\
\multicolumn{1}{l}{Mar. 6, 1991} & NTT & $B,V,R$ & 1,1,1 & 0.1,0.25 \\
\multicolumn{1}{l}{Mar. 12, 1991} & Danish & $B,V$ & 3,3 & 0.66,1.33 \\
\multicolumn{1}{l}{Mar. 13, 1991} & Danish & $B,V$ & 2,2 & 0.66,1.33 \\
\multicolumn{1}{l}{Mar. 14, 1991} & Danish & $B,V$ & 3,3 & 0.66,1.33 \\
\multicolumn{1}{l}{Mar. 16, 1991} & Danish & $B,V$ & 1,4 & 0.66,1.33 \\
\multicolumn{1}{l}{Mar. 17, 1991} & Danish & $B,V$ & 2,2 & 0.66,1.33 \\
\multicolumn{1}{l}{Mar. 18, 1991} & Danish & $B,V$ & 2,2 & 0.66,1.33 \\
\multicolumn{1}{l}{Mar. 25, 1991} & NTT & $B,V,R$ & 1,1,1 & 0.16,0.5 \\
\multicolumn{1}{l}{May 19, 1991} & 3.6m & $B,V,R$ & 1,1,1 & 0.5,1.5 \\
\multicolumn{1}{l}{Jan. 1, 1992} & NTT & $B,V,R$ & 1,1,1 & 1,2,3 \\
\noalign{\smallskip}
\hline
\noalign{\smallskip}
\hline
\noalign{\smallskip}
\multicolumn{5}{c}{Spectra} \\
\noalign{\smallskip}
\hline
\noalign{\smallskip}
\multicolumn{1}{l}{Jan. 15, 1991} & NTT & \#3 & 2 & 3 \\
\multicolumn{1}{l}{Jan. 18, 1991} & 1.5m & $Gr$15,$Gr$17 & 1,1 & 15,18.3 \\
\multicolumn{1}{l}{Jan. 19, 1991} & 1.5m & $Gr$15,$Gr$17 & 1,1 & 25,40 \\
\multicolumn{1}{l}{Jan. 20, 1991} & 1.5m & $Gr$17 & 1 & 50 \\
\multicolumn{1}{l}{Jan. 23, 1991} & NTT & \#5 & 1 & 10 \\
\multicolumn{1}{l}{Feb. 15, 1991} & 1.5m & 3750--7500 \AA~ & 1 & 20 \\
\multicolumn{1}{l}{Feb. 21, 1991} & 1.5m & 3500--10100 \AA~ & 1 & 15 \\
\multicolumn{1}{l}{Mar. 6, 1991} & NTT & \#2,$Gr$5 & 1,1 & 10,30 \\
\multicolumn{1}{l}{Mar. 21, 1991} & 2.2m & \#7 & 1 & 30 \\
\multicolumn{1}{l}{Mar. 25, 1991} & NTT & \#5,\#6 & 2,1 & 10,20 \\
\multicolumn{1}{l}{May 5, 1991} & 1.5m & $Gr$21 & 1 & 20 \\
\multicolumn{1}{l}{May 19, 1991} & 3.6m & $B$300,$R$300 & 1,1 & 10,15 \\
\noalign{\smallskip}
\hline
\end{tabular}
\end{center}
\end{table*}

X--ray and optical observations during the decay showed two important 
behaviours: the presence of an electron--positron annihilation line at 0.511 
MeV (Sunyaev et al. 1992, Goldwurm et al. 1992), which is considered a clue 
for the presence of a black hole (see the review by Tanaka \& Lewin 1995), and
the appearance of superhumps (Bailyn 1992).
The latter phenomenon, first seen in SU UMa--type DNe, takes place only if the 
mass ratio $q=M_2/M_1$ is less than 0.25--0.33 (Whitehurst \& King 1991).
Since the mass of the secondary in SXTs is $\ga$0.5 $M_\odot$ (van Paradijs \&
McClintock 1995, Tanaka \& Lewin 1995), the accretor should be a 
highly--collapsed object. Actually, superhumps in SXTs have been observed in 
outbursting QZ Vul (Charles et al. 1991), V518 Per (Kato et al. 1995), V2293 
Oph (Masetti et al. 1996) and MM Vel (Masetti et al. 1997; Della Valle et al. 
1997).

The X--ray lightcurve also showed two secondary maxima at $\sim$70
(Kitamoto et al. 1992) and $\sim$200 days (Ebisawa et al. 1994) after the main
X--ray peak. On the contrary, until now there is no indication of optical 
secondary maxima or minioutbursts.

GU Mus reverted to quiescence about one year after the outburst (Della Valle 
1992). In April 1992, when the star was already at minimum, Remillard et al. 
(1992), found that the mass function of the primary is $\sim$3 
$M_\odot$, which is already beyond the maximum allowed mass for a neutron star 
(Rhodes \& Ruffini 1974). This value placed GU Mus amongst the galactic 
black--hole candidates. 

In this paper we present the spectrophotometric follow--up of GU Mus, from the
discovery to the late decline. Section 2 illustrates the observations and the 
reduction of images and spectra. Section 3 shows the analysis of the data, and 
Sect. 4 discusses the results and draws the conclusions.

\section{Observations and data reduction}

All the images and the spectra were taken at La Silla with different ESO 
telescopes. Table 1 gives the complete journal of observations.
The upper part of Table 1 lists the images, while the lower part reports the 
spectroscopic observations.

We have globally obtained 128 images (53 in the $B$, 41 in the $V$ and 34 in the
$R$ Johnson bands) and 19 spectra (with slit widths varying from 1" to 2",
giving a resolution between 2.0 and 9.6 \AA/pixel) over a period of nearly one 
year (from January 15, 1991 to January 1, 1992). However most observations 
are concentrated within the first three months after the outburst.
Standard bias and flat field corrections were performed on each frame. 

Images were reduced with DAOPHOT II (Stetson 1987) and {\sl ALLSTAR} inside 
MIDAS. The nova has been calibrated in $B$ and $V$ magnitudes by using the 
field stars 1, 2 and 3 (Bailyn 1992).
We have measured the $R$ magnitudes of these three stars through Mark A and T 
Phe Landolt fields (Landolt 1992), and obtained $R=13.20 \pm 0.02$, $R=14.49 
\pm 0.02$ and $R=14.54 \pm 0.02$, respectively.
Table 2 reports the $B$, $V$ and $R$ magnitudes of the object, together with 
the ($B-V$) and ($V-R$) color indexes. The typical 
photometric error for each measurement is of the order of 
$\pm$0.05 mag in the $B$ band and $\pm$0.03 mag in both $V$ and $R$.

\begin{table*}
\caption[]{$B$, $V$ and $R$ magnitudes, along with the ($B-V$) and ($V-R$) 
colors, of GU 
Mus. The typical error is $\pm$0.05 mag for the $B$ values and $\pm$0.03 mag 
for $V$ and $R$. When more than one measurement in the same band was available 
during the same night, the mean value has been reported. The magnitudes are 
referred to the mean Heliocentric Julian Day (HJD) of the observations}
\begin{center}
\begin{tabular}{cccccc}
\noalign{\smallskip}
\hline
\noalign{\smallskip}
 HJD & $B$ & $V$ & $R$ & ($B-V$) & ($V-R$) \\
\noalign{\smallskip}
\hline
\noalign{\smallskip}
 2448271.73 & 13.66 & 13.43 & 13.24 & 0.23 & 0.19 \\
 2448272.72 & 13.84 & 13.60 & 13.33 & 0.24 & 0.27 \\
 2448273.68 & 13.97 & 13.68 & 13.42 & 0.29 & 0.26 \\
 2448274.84 & 13.91 & 13.62 & 13.36 & 0.29 & 0.26 \\
 2448275.86 & 14.07 & 13.77 & 13.57 & 0.30 & 0.20 \\
 2448276.86 & 13.94 & 13.67 & 13.41 & 0.27 & 0.26 \\
 2448277.87 & 13.97 & 13.71 & 13.46 & 0.28 & 0.25 \\
 2448278.85 & 13.92 & 13.63 & 13.43 & 0.29 & 0.20 \\
 2448279.80 & --- & 13.64 & 13.41 & --- & 0.23 \\
 2448280.75 & --- & 13.80 & 13.46 & --- & 0.24 \\
 2448285.83 & 14.05 & 13.75 & 13.51 & 0.30 & 0.24 \\
 2448298.87 & --- & --- & 13.60 & --- & --- \\
 2448304.78 & 14.38 & 14.13 & 13.88 & 0.25 & 0.25 \\
 2448305.77 & 14.39 & 14.12 & 13.87 & 0.27 & 0.25 \\
 2448306.90 & 14.57 & 14.29 & 14.04 & 0.28 & 0.25 \\
 2448308.85 & 14.60 & 14.23 & 13.95 & 0.37 & 0.28 \\
 2448311.77 & 14.60 & 14.30 & 14.12 & 0.30 & 0.18 \\
 2448315.75 & 14.46 & 14.30 & 14.06 & 0.16 & 0.24 \\
 2448316.78 & 14.45 & 14.31 & 14.08 & 0.14 & 0.23 \\
 2448317.80 & 14.49 & 14.37 & 14.10 & 0.12 & 0.27 \\
 2448321.63 & 14.66 & 14.46 & 14.20 & 0.20 & 0.26 \\
 2448327.74 & 14.78 & 14.45 & --- & 0.33 & --- \\
 2448328.73 & 14.71 & 14.36 & --- & 0.35 & --- \\
 2448329.74 & 14.83 & 14.47 & --- & 0.36 & --- \\
 2448331.75 & 14.80 & 14.45 & --- & 0.35 & --- \\
 2448332.70 & 14.69 & 14.34 & --- & 0.35 & --- \\
 2448333.71 & 14.88 & 14.53 & --- & 0.35 & --- \\
 2448340.73 & 14.69 & 14.48 & 14.21 & 0.21 & 0.27 \\
 2448395.51 & 15.81 & 15.40 & 15.12 & 0.41 & 0.28 \\
 2448622.85 & 22.1$\pm$0.3 & 20.5$\pm$0.1 & 19.5$\pm$0.1 & 1.6$\pm$0.3 & 
1.0$\pm$0.15 \\
\noalign{\smallskip}
\hline
\end{tabular}
\end{center}
\end{table*}

The reduction of the spectroscopic material was carried out with IRAF. He--Ar 
lamps were used for the wavelength calibration, and the standards Feige 56, 
EG 76, L970--30, LTT 3218, Hiltner 600, CD--32, Wolf 485A and EG 274 for the 
flux calibration. We corrected the spectra for interstellar reddening 
with the prescription of Cardelli et al. (1989), and using $E(B-V)=0.30$ as 
derived by Della Valle et al. (1991).

Finally, the observation times have been corrected to heliocentric times of 
mid--exposure.

\section{Data analysis}

\subsection{Photometry}

The lightcurves in $B$, $V$ and $R$ bands are shown in Fig. 1a,b,c.
Light fluctuations are seen in each lightcurve, with
amplitudes up to 0.2 mag. These light variations are erratic in the early 
decline, and later they appear as small light oscillations with typical 
timescales of about 10 days (see also Fig. 5a).

Figure 2 reports the X--ray and the $B$ lightcurves in a common flux scale.
The comparison between the X--ray peak flux (Kitamoto et al. 1992) and the 
optical flux computed from the spectra at maximum light (see Fig. 3) leads to a 
ratio of $\sim$10$^{3}$, which is typical for LMXBs and outbursting SXTs 
(Tanaka \& Lewin 1995).

\begin{figure}
\psfig{file=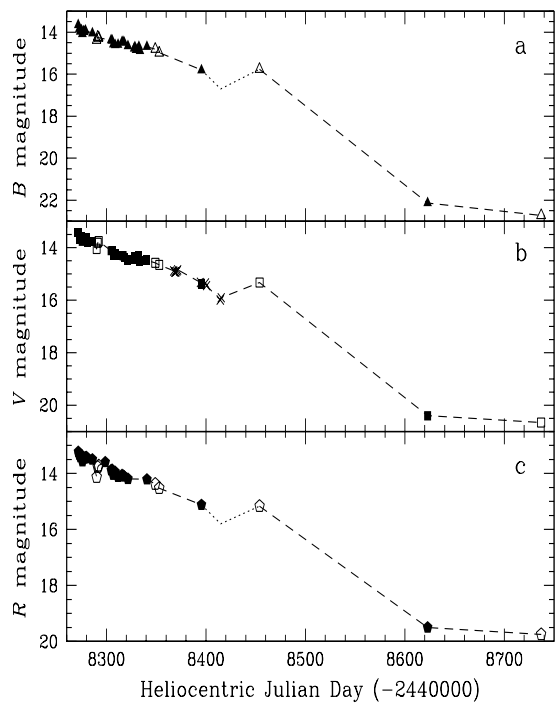,height=11cm,width=8.5cm,angle=0}
\caption[]{{\bf a} $B$, {\bf b} $V$ and {\bf c} $R$ lightcurves of
the outburst of GU Mus. We indicate, with a different hyphenation, the 
possible presence in the $B$ and $R$ bands of the secondary maximum seen in 
the $V$ lightcurve $\sim$200 days after the X--ray peak. Open symbols indicate 
the magnitude determinations made by King et al. (1996), while crosses 
correspond to the nightly means of the observations made by Bailyn (1992)}
\end{figure}

\begin{figure}
\psfig{file=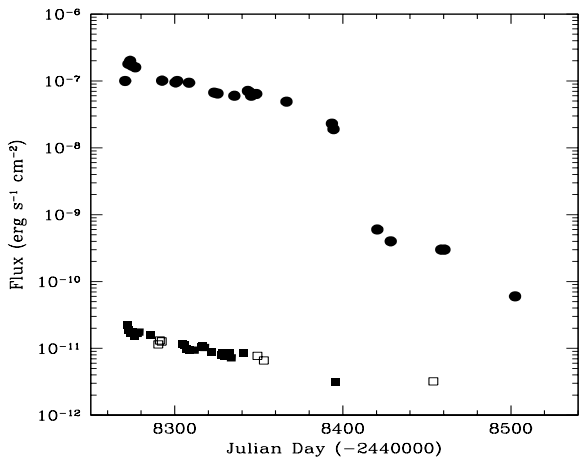,height=7cm,width=8.5cm,angle=0}
\caption[]{X--ray (circles) and $B$ (squares) lightcurves of the outburst of 
GU Mus. X--ray data are extracted from Table 2 of Ebisawa et al. (1994). Open 
squares indicate the $B$ magnitude determinations made by King et al. (1996). 
Note their similar trends and the coincidence of their secondary maxima}
\end{figure}

A linear fit on the decay lightcurves has been performed to estimate 
the decline rates in the $B$, $V$ and $R$ bands. Table 3 reports the decay 
rates for each photometric band in different parts of the lightcurve.

\begin{table}
\caption[]{Decay rates (in mag d$^{-1}$) of the $B$, $V$ and $R$ outburst 
lightcurves of GU Mus computed at different phases of the decline}
\begin{center}
\begin{tabular}{cccc}
\noalign{\smallskip}
\hline
\noalign{\smallskip}
Decline phase & $B$ & $V$ & $R$ \\
\noalign{\smallskip}
\hline
\noalign{\smallskip}
\multicolumn{1}{l}{January 1991} & 0.0199 & 0.0161 & 0.0135 \\
\multicolumn{1}{l}{Feb.--Mar. 1991} & 0.0109 & 0.0091 & 0.0090 \\
\multicolumn{1}{l}{Apr.--May 1991} & 0.0203 & 0.0168 & 0.0166 \\
\multicolumn{1}{l}{May 1991--Jan. 1992} & 0.0278 & 0.0220 & 0.0193 \\
\multicolumn{1}{l}{Jan. 1991--Jan. 1992} & 0.0241 & 0.0199 & 0.0178 \\
\noalign{\smallskip}
\hline
\end{tabular}
\end{center}
\end{table}

Two bumps
at $\sim$50 and $\sim$70 days after the maximum are clearly visible
(see also Fig. 5a). The second bump is 
consistent with the secondary maximum observed in the X--ray band (see Fig. 2) 
and reported by Kitamoto et al. (1992), while the first one might be associated
to a small feature in the X--ray lightcurve (see Fig. 1, upper panel, of 
Kitamoto et al. 1992; see also Fig. 4 of Greiner et al. 1994). This behaviour 
is typical of SXTs during the decline (Chen et al. 1993, van Paradijs \& 
McClintock 1995).

The decay starts again in the April--May period, and strengthens
in the second part of 1991. It can be also noticed from Table 3 that, after 
March 1991, the decay in the $B$ is slightly faster than the ones in $V$ and 
$R$. 

However, if we integrate our photometric data with the magnitude determinations 
reported by King et al. (1996) and with the nightly averages of the Bailyn's 
(1992) observations, we notice another local maximum in the $V$ lightcurve 
around Julian Day (hereafter JD) 2448460, i.e. $\sim$200 days after the X--ray
peak: thus, the increased luminosity can be correlated with the tertiary 
X--ray maximum (see Fig. 2) observed by Ebisawa et al. (1994). 

The presence of minioutbursts during the late decline of
MM Vel (=X--Ray Nova Velorum 1993; Bailyn \& Orosz 1995) and
V518 Per (=GRO J0422+32; Chevalier \& Ilovaisky 1995, Callanan et al. 1995),
would indicate that they are a not unusual feature in these objects, though 
for these two
systems this phenomenon has been purely optical and not associated with a
restart of the X--ray activity.

The magnitudes at minimum, as measured on January 1, 1992 
(see Table 2), are in agreement with the mean magnitude 
determinations by Remillard et al. (1992) and by King et al. (1996).

\begin{figure}
\psfig{file=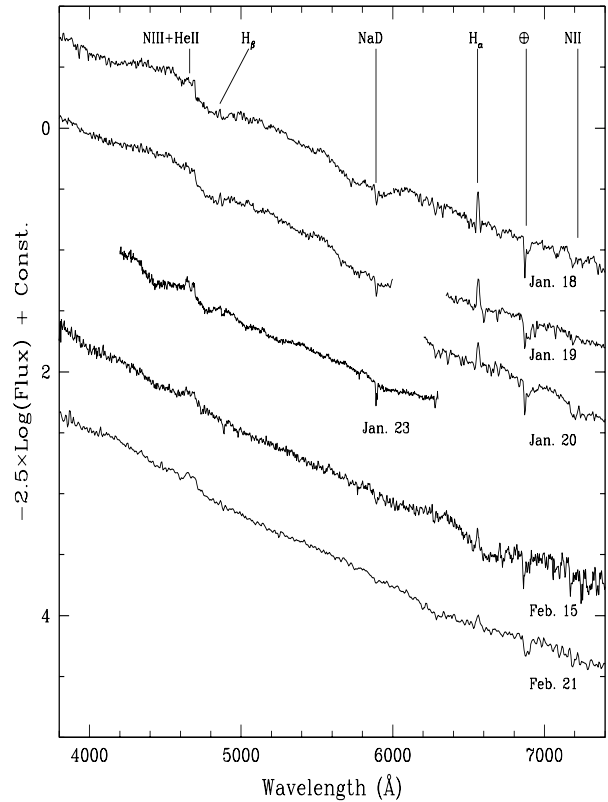,height=11cm,width=8.5cm,angle=0}
\caption[]{Six spectra of GU Mus acquired between January and February 1991 and
listed in Table 1. The main spectroscopic features are indicated.
Fluxes are in units of erg s$^{-1}$ cm$^{-2}$. For sake of clarity, the 
spectra have been separated by 0.5 logarithmic flux units. The additive 
constant for the spectrum at the top of the figure is $-$33.5}
\end{figure}

\begin{figure}
\psfig{file=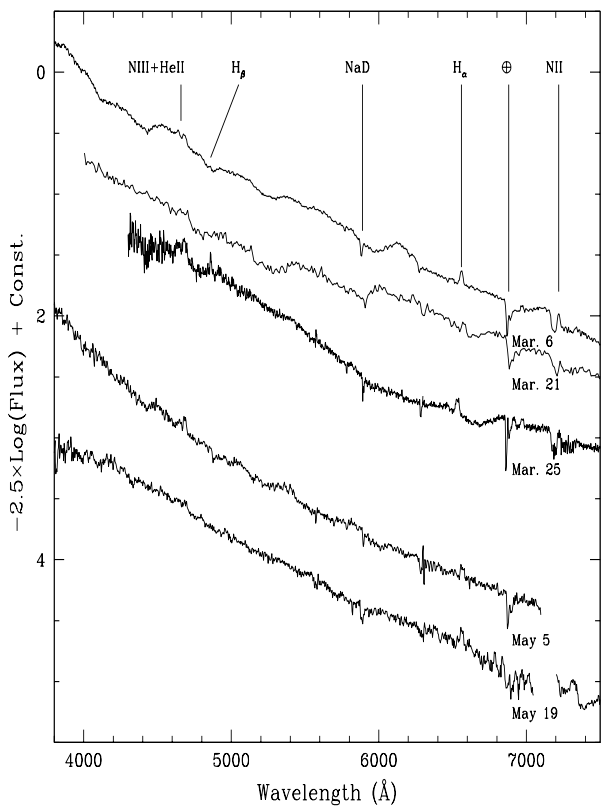,height=11cm,width=8.5cm,angle=0}
\caption[]{Five spectra of GU Mus taken between March and May 1991 and listed 
in Table 1. The main spectroscopic features are indicated. Fluxes are in units 
of erg s$^{-1}$ cm$^{-2}$. For sake of clarity, the spectra have been 
separated by 0.5 logarithmic flux units. The additive constant for the spectrum
at the top of the figure is $-$33.5}
\end{figure}

\medskip
The evolution of ($B-V$) and ($V-R$) colors during the decay is shown in 
Table 2. It is noteworthy that, while the ($V-R$) color is roughly constant
around 0.25 from the beginning of the outburst until mid--May 1991, the 
($B-V$) seems to decay (0.00118 mag d$^{-1}$) as the decline proceeds.
The mean ($B-V$) color fairly agrees with the value given by
Bailyn (1992) for the month of April 1991.
We also note that the object appears to become bluer around JD 
2448320 and $\sim$20--25 days later, that is during the two bumps in the
early $B$ lightcurve (see Fig. 5a).

At the beginning of January 1992, the colors return to their pre--outburst 
values (see Della Valle et al. 1991).

\subsection{Spectroscopy}

The spectra taken during the follow--up period are listed in Table 1 and
shown in Figs. 3 and 4.

The most prominent emission lines are H$_\alpha$, H$_\beta$,
the N {\sc iii}+He {\sc ii} blend at 4640--4686 \AA~and, from
March 1991 (Fig. 4; first spectrum from top), N {\sc ii} $\lambda$7217. 
The fluxes of these emission lines are reported in Table 4. The only 
evident absorption lines are the interstellar NaD doublet at 5890 \AA~and the 
telluric lines at 6850 \AA~ and 7600 \AA~ (the latter has not been included in 
the figures).

\begin{table*}
\caption[]{Fluxes (in units of 10$^{-15}$ erg s$^{-1}$ cm$^{-2}$) of the main 
emission features seen in the spectra of Figs. 3 and 4.}
\begin{center}
\begin{tabular}{cccccc}
\noalign{\smallskip}
\hline
\noalign{\smallskip}
Decline phase & H$_\alpha$ & H$_\beta$ & He {\sc ii} & N {\sc iii} 
& N {\sc ii} \\
\noalign{\smallskip}
\hline
\noalign{\smallskip}
\multicolumn{1}{l}{Jan. 18} & $78\pm8$ & $43\pm4$ & $99\pm9$ & $180\pm20$ 
& --- \\
\multicolumn{1}{l}{Jan. 19} & $71\pm7$ & $46\pm5$ & $90\pm8$ & $150\pm15$ 
& --- \\
\multicolumn{1}{l}{Jan. 20} & $55\pm6$ & (abs.) & --- & --- & $18\pm2$ \\
\multicolumn{1}{l}{Jan. 23} & --- & (abs.) & $94\pm9$ & $170\pm20$ & --- \\
\multicolumn{1}{l}{Feb. 15} & $56\pm6$ & --- & $74\pm7$ & $120\pm10$ & --- \\
\multicolumn{1}{l}{Feb. 21} & $20\pm3$ & --- & $58\pm9$ & $75\pm10$ & --- \\
\multicolumn{1}{l}{Mar. 6} & $18\pm3$ & $9\pm2$ & $37\pm6$ & $48\pm7$ & 
$16\pm3$ \\
\multicolumn{1}{l}{Mar. 21} & $28\pm4$ & --- & $36\pm5$ & $54\pm8$ & --- \\
\multicolumn{1}{l}{Mar. 25} & $24\pm4$ & $46\pm7$ & $160\pm30$ & $83\pm10$ & 
$16\pm3$\\
\multicolumn{1}{l}{May 5} & $8\pm2$ & --- & $27\pm5$ & $13\pm3$ & --- \\
\multicolumn{1}{l}{May 19} & $10\pm2$ & --- & $16\pm3$ & $12\pm3$ & --- \\
\noalign{\smallskip}
\hline
\end{tabular}
\end{center}
\end{table*}

The mean value of the EW of the NaD absorption is $1.4\pm0.1$ \AA,
in good agreement with the estimate of Della Valle et al. (1991). 
According to the relation between the EW of NaD and the ($B-V$) color excess 
given by Barbon et al. (1990), we find
($B-V$)$_{\rm o}$ $\sim$ +1.3 at quiescence, consistent with the spectral 
classification by Remillard et al. (1992) and by Orosz et al. (1996).

Figure 5 shows the first part of the $B$ lightcurve decline together with EW's
and FWHM's of the main emission lines visible in the spectra of Figs. 3 and 4.

The spectral evolution of the nova during the decline is mainly characterized by
the weakening and the broadening of the H$_\alpha$ emission, as shown in Fig. 
5b,c. An increase of the FWHM's is actually observed also in the He {\sc ii} 
$\lambda$4686 emission, while the behaviour of the $\lambda$4640 blend is more 
uncertain due to the broadened profile and to the bad signal--to--noise 
ratio (see Fig. 5c).
On February 21 and on March 21, H$_\alpha$ seems to be embedded inside a 
wide and shallow absorption. This is not unusual for SXTs (see e.g. Callanan et
al. 1995, Masetti et al. 1997, Bianchini et al. 1997).

The profile of H$_\beta$, close to maximum light, shows a shallow absorption 
filled in by an emission core. During the secondary X--ray maximum 
(Fig. 4; third spectrum from top) only the emission component is shown.

The He {\sc ii} and $\lambda$4640 blend fluxes decay monotonically with time 
but rise up during the X--ray `reflare'. At that time (see Fig. 5b,c) also 
the EW and FWHM of He {\sc ii} show a ``jump".

The N {\sc ii} is clearly present only in the March spectra (in Fig. 4), and 
perhaps in the spectrum of January 20 (Fig. 3; third spectrum from top), 
with an average EW of $\sim$2 \AA.

The continuum becomes less steep in the blue part as the outburst proceeds;
actually, the last spectrum of the run appears flatter than the other ones. At 
this stage, the disk emission is still predominant, indeed we do not see any 
absorption feature of the secondary, since the magnitude of the system is still
around $V\sim15.5$ and the secondary is a K3--K5 dwarf (Orosz et al. 1996).

\section{Discussion}
\subsection{The lightcurves}

We have detected a bump in the $B$ and ($B-V$) lightcurves very close to the 
secondary X--ray maximum (see Fig. 5a; see also Fig. 2), and a 
brightening in the $V$ band (and possibly also in $B$ and $R$) close to the 
tertiary X--ray maximum at $\sim$200 days after the peak.

It is noteworthy that these minioutbursts are different from those shown by
MM Vel (Bailyn \& Orosz 1995) and by V518 Per (Chevalier \& Ilovaisky 1995) 
because the latter objects show purely optical `reflares' 
well after the end of the X--ray outburst which have no X--ray counterpart. 

From Table 3 we also note that the decay is slower as we move to higher 
wavelengths: this may be due to the cooling of the X--ray illuminated zones of 
the binary system (i.e., the outer disk and the inner face of the secondary).

The light fluctuations observed during the first decline (Fig. 5a) might be due
to a superhump activity which could have been present also before it was 
observed by Bailyn (1992). 

Other longer--term light fluctuations, observed during the decline, might be 
real and correspond to faint secondary maxima. In particular, a sort of 10--day
periodicity appears to be present during the decline in the $B$ lightcurve 
(Fig. 5a). According to Warner (1995; and references therein), this behaviour 
is similar to that shown by some classical novae during the transition phase.

\subsection{The spectra and the disk stability}

During the first five days of the decline the EW of H$_\alpha$ drops quite 
steeply by a factor 2.5, then it remains at about the same value throughout 
the first half of 1991.

\begin{figure}
\psfig{file=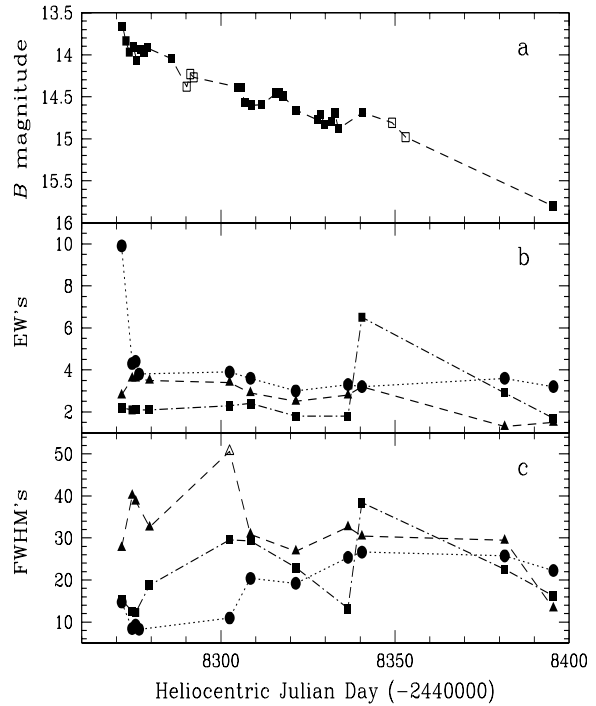,height=11cm,width=8.5cm,angle=0}
\caption[]{Evolution of the $B$ magnitude (a), EW's (b) and FWHM's (c) of 
H$_\alpha$ (circles), He {\sc ii} $\lambda$4686 (squares) and 
N {\sc iii} $\lambda$4640 blend (triangles) between January and May 1991. 
EW's and FWHM's are both given in \AA. The empty triangle in the lower panel 
corresponds to a rather uncertain measurement, while the open squares in the 
upper panel indicate the observations of King et al. (1996)}
\end{figure}

\begin{figure}
\psfig{file=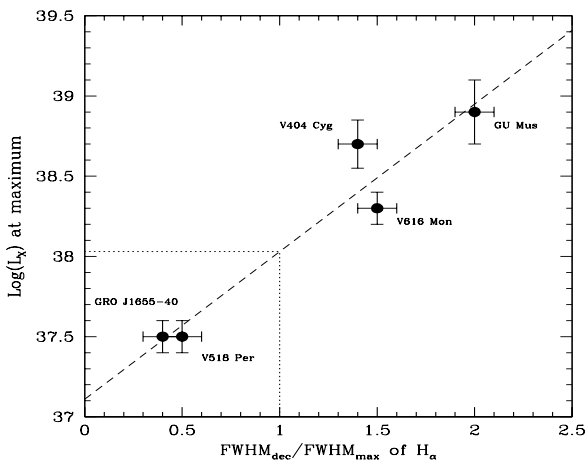,height=7cm,width=8.5cm,angle=0}
\caption[]{Correlation between the X--ray luminosity (in erg s$^{-1}$) 
at maximum and the ratio of FWHM's of H$_\alpha$ emission at 
mid--decline and at the beginning of the outburst for 5 SXTs. 
The long--dashed line represents the least--squares fit of the data points}
\end{figure}

\begin{table*}
\caption[]{Optical decline rates, distances, times $\Delta$t from the light 
maximum at which the FWHM$_{\rm dec}$ of H$_\alpha$ was measured and 
corresponding magnitude drops ($\Delta$m) for the 5 SXTs quoted in Sect. 4.2}
\begin{center}
\begin{tabular}{cccccc}
\noalign{\smallskip}
\hline
\noalign{\smallskip}
Object & Optical decline & Distance & $\Delta$t & $\Delta$m & Ref. No. \\
 & rate (mag d$^{-1}$) & (kpc) & & & \\
\noalign{\smallskip}
\hline
\noalign{\smallskip}
\multicolumn{1}{l}{V518 Per} & 0.009 & 2 & $\sim$200 & $\sim$2 & 1,2,3 \\
\multicolumn{1}{l}{V616 Mon} & 0.015 & 1.05 & $\sim$200 & $\sim$3 & 4,5 \\
\multicolumn{1}{l}{GU Mus} & 0.016 & 5.5 & $\sim$150 & $\sim$2.5 & 6,7 \\
\multicolumn{1}{l}{GRO J1655-40} & 0.016 & 3.2 & $\sim$200 & $\sim$3 & 
8,9,10,11 \\
\multicolumn{1}{l}{V404 Cyg} & 0.009 & 2.6 & $\sim$180 & $\sim$2 & 8,12 \\
\noalign{\smallskip}
\hline
\noalign{\smallskip}
\multicolumn{6}{l}{References. --- 1. Castro--Tirado et al. 1993;
2. Chevalier \& Ilovaisky 1995; 3. Shrader et al. 1994;} \\
\multicolumn{6}{l}{4. Whelan et al. 1977; 5. Shahbaz et al. 1994; 6. this work;
7. Orosz et al. 1996;} \\
\multicolumn{6}{l}{8. Goranskij et al. 1996; 9. Hjellming \& Rupen 1995; 
10. Bailyn et al. 1995a;} \\
\multicolumn{6}{l}{11. Bianchini et al. 1997; 12. Casares et al. 1993.} \\
\noalign{\smallskip}
\hline
\end{tabular}
\end{center}
\end{table*}

The He {\sc ii} line shows a bump which approximately starts at JD 2448340,
in coincidence with the secondary maximum of the 
X--ray (Kitamoto et al. 1992) and the $B$ lightcurves.
The increase of luminosity in the high--energy bands (UV and X--rays) 
should then be responsible for this bump, since the strength of this 
emission line is actually correlated with the UV continuum level 
(Garnett et al. 1991).
Later on, the He {\sc ii} line becomes weaker, as also shown by Cheng et al. 
(1992; their Fig. 3c).

The EW of the $\lambda$4640 blend appears slightly stronger than 
that of He {\sc ii} before JD 2448340 and it
seems to remain at about the same level throughout the brightening of the 
He {\sc ii} component. It shows no increase when the X--ray and the $B$ 
lightcurves have a secondary maximum (see Fig. 2 and Fig. 5a,b).
This behaviour could be explained by the rather broadened and undefined 
profile of the $\lambda$4640 blend and/or by its lower excitation potential 
with respect to that of He {\sc ii}.

The FWHM (Fig. 5c) of H$_\alpha$ and of He {\sc ii} tend to increase 
with time. 
The FWHM of the $\lambda$4640 blend starts with larger values, and then it 
seems to decrease. The widening at JD 2448304 shown in Fig. 5c is rather 
uncertain due to the difficult determination of the profile.
The FWHM's of He {\sc ii} and of the $\lambda$4640 blend seem to 
decrease beginning on May 1991. Anyway, the lack of spectroscopic 
observations after May 19, 1991 does not allow us to fully confirm this trend.
This fact would indicate that during the decline the emission region 
moves inward in the disk, that is, towards larger keplerian velocity radii. 

However we note that X--ray novae may present opposite 
behaviours. Some objects, like V518 Per (Shrader et al. 1994) 
and GRO J1655-40 (Bianchini et al. 1997), show emission lines
with decreasing widths with the time; some other ones, like GU Mus (this work),
V404 Cyg (Gotthelf et al. 1992) and V616 Mon (Whelan et al. 1977) 
show emission lines which become larger and larger during the decline. 

To investigate this point we plot in Fig. 6 the ratio 
FWHM$_{\rm dec}$/FWHM$_{\rm max}$ of the H$_\alpha$ line at about 150--200 days
after the X--ray peak and at maximum versus the X--ray luminosity at maximum 
of the 5 SXTs with available data (in Table 5), i.e. GU Mus
(this work), V404 Cyg (Gotthelf et al. 1992), V616 Mon (Whelan et al. 1977), 
GRO J1655-40 (Bianchini et al. 1997) and V518 Per (Shrader et al. 1994). 
The errors on Log({\it L$_{\rm X}$}) are mainly due to the uncertainty on the 
distance of the objects (see Table 5),
while those affecting FWHM ratios are originated by the signal--to--noise 
ratio of the spectra. The long--dashed line represents the weighted 
least--squares fit of the data points:

\begin{equation}
{\rm Log}(L_{\rm X}) = 0.94 (\pm 0.15) {FWHM_{\rm dec} \over 
	FWHM_{\rm max}} + 37.1 (\pm 0.2).
\end{equation}

The correlation coefficient of the data is 0.96, which indicates
(admittedly on the basis of a scanty statistic), that these quantities seem
to be correlated. 

The regression line of Fig. 6 suggests a peak luminosity $\ga$$10^{38}$ erg 
s$^{-1}$ for the case FWHM$_{\rm dec}$/FWHM$_{\rm max}$=1.
This value of the luminosity approximately represents 
the Eddington luminosity of SXTs. 
We note that the behaviour of the sub--Eddington objects would typically 
represent the one it is expected from the outburst caused by an enhancement 
of mass transfer from the secondary. Actually, former disk instability event 
has only triggered the mass transfer paroxism.
In fact, during the early outburst the disk becomes smaller because of 
the very large flow of low angular momentum material from the X--ray 
heated secondary; so we may observe larger emission lines. 
During the decline the disk relaxes and the emission region will 
expand again towards lower keplerian velocities 
as it was suggested for GRO J1655-40 by Bianchini et al. (1997).

For super--Eddington objects, like GU Mus, accretion from the inner regions of
the disk should be inhibited and the disk itself might be partially disrupted.
More likely, we might argue that at super--Eddington accretion luminosities 
the disk turns its structure into a `geometrically thick' one (see e.g. Frank 
et al. 1992), in which only the outer cooler regions can be seen at almost 
all inclinations. 
In this case, at light maximum the line emitting region is placed at quite 
large radii, thus producing narrower emission lines than during the decline, 
when a `standard' disk is formed again.

\begin{acknowledgements}

We are indebted to the anonymous referee for stimulating comments. We also 
thank R. Canterna for his useful suggestions.

\end{acknowledgements}

\end{document}